\documentclass[prb,aps,amssymb,twocolumn,showpacs,superscriptaddress,footinbib,10pt]{revtex4-1}
\usepackage{graphicx}
\usepackage{amsmath,amsfonts,amssymb}
\usepackage{xcolor}
\usepackage{hyperref}
\usepackage{upgreek}
\usepackage{float}

\begin{document}
\title{Role of optical rectification in photon-assisted tunneling current}%

\author{P. F\'{e}vrier, J. Basset, J. Estève, M. Aprili, J. Gabelli}%
\affiliation{Laboratoire de Physique des Solides, CNRS, Universit\'{e} Paris-Sud, Universit\'{e} Paris-Saclay, Orsay, France}

\date{}

\begin{abstract}
We study the optical rectification in a metallic tunnel junction. We consider a planar  junction in a Kretschmann configuration and measure the photon-assisted tunneling under infrared illumination at $\lambda= 1.5\, \mu \mathrm{m}$. To address the microscopic mechanism at the origin of the optical rectification, we compare the photon assisted current and the  current-voltage characteristics of the junction measured on a voltage range much greater than $V_0=\frac{hc}{e\lambda}=0.825 \, \mathrm{V}$.  The experimental results do not agree with the Tucker theory based on the exchange of energy quanta between electrons and photons and describing the dc current induced by photon-assisted processes in terms of a linear combination of the shifted characteristics $I(V)$ and $I(V\pm V_0)$. We show instead that the illumination power mainly goes into heating and that the rectification results mainly from the non-linearity of the tunnel junction at optical frequency.
\end{abstract}

\maketitle
\section{Introduction}

\noindent  Recent years have seen tremendous progress in nanophotonics with the realization of optical antennas which can receive \cite{Muhlschlegel2005,Bharadwaj2009} and emit radiation at optical frequencies \cite{Qian_2018,Zhang_2019,Qian_2021}. In terms of detectors, nanoantennas allow an efficient transfer of  radiation into surface plasmons and they close the gap between the characteristic lengths of optics (micrometer scale) and electronic devices (nanometer scale). Thus, if coupled with a non-linear element, they can be used for optical rectification. One thus refers to rectenna: a nanoantenna  converts the optical radiation into an ac voltage which is in turn converted to a dc current through a rectifier (diode). However, optical rectification in conductors raises central questions. (i) Is it possible to define a voltage at optical frequencies as used in the microwave domain? A microwave excitation of frequency $\omega$ at low temperature ($k_{\mathrm{B}}T \ll \hbar \omega$) actually induces discontinuities in the derivative of distribution function at energies $\epsilon = n \hbar \omega$ with $n$ integer \cite{Kozhevnikov_2000}. (ii) Is it possible to observe this kind of step like distribution function at optical energies?  While they are linear elements at low voltage bias (i.e $V_\mathrm{dc} \leq 10 \, \mathrm{mV}$ ), Metal-Insulator-Metal (MIM) junction  becomes strongly non-linear at optical frequencies ($\hbar \omega \sim 1 \, \mathrm{eV} $) and are  promising rectifiers. The goal of this article is to focus on MIM junctions as an optical rectifier and to address the underlying microscopic mechanism.

\noindent The photon-assisted tunnelling (PAT) has been investigated in experiments using microwave radiation, where the illumination is simply described by an ac bias voltage, $V_{\mathrm{ac}} \cos \omega t$, across the conductor \cite{Tucker_1985,Kouwenhoven_1994,Gabelli_2008}. In this regime, the PAT is well described in the quantum limit, $\hbar \omega \gg k_{\mathrm{B}} T$, by the approach of Tien and Gordon \cite{Tien_1963} where this time dependent potential, creates a non-equilibrium distribution function in the reservoir absorbing the radiation and corresponding to energy quanta exchange between electrons and microwave photons. It is tempting to describe the optical rectification within the same theory by describing the illumination by an optical voltage $V_{\mathrm{ac}}$. In the following, we consider the excess current due to illumination: $\delta I(V_{\mathrm{dc}},V_{\mathrm{ac}})=I(V_{\mathrm{dc}},V_{\mathrm{ac}})-I_0(V_{\mathrm{dc}})$  where $I_0$ is the current in the absence of incident light and $I(V_{\mathrm{dc}},V_{\mathrm{ac}})$ the current under illumination.  As demonstrated below , the time-averaged photon-assisted tunneling current calculated from the distribution functions  predicted by Tien and Gordon, is then given at the first non-vanishing order in $V_{\mathrm{ac}}$, by the Tucker formula (TF):

\begin{equation}\label{eq.Tucker}
\begin{split}
\delta I(V_{\mathrm{dc}},V_{\mathrm{ac}})=& \left( \frac{eV_{ac}}{2\hbar \omega}\right)^2 \left[I_0\left( V_{\mathrm{dc}}+V_0 \right)  \right.\\
& \left.+I_0\left( V_{\mathrm{dc}}-V_0 \right)-2I_0\left(V_{\mathrm{dc}}\right)\right],
\end{split}
\end{equation}

\noindent where $V_0=\hbar \omega/e$, $\hbar$  is the Planck constant and $e$ the elementary charge. Below we will address the full quantum regime  $V_0 \approx V_{\mathrm{dc}}$.  This  is technically demanding as it requires to be able to polarize a MIM junction with a voltage bias  $V_{\mathrm{dc}}>V_0 \approx 1 V$ without deteriorating the device. We are not aware of previous work in this regime. To do so, we have chosen to perform the experiment on a robust Al/AlOx/Al tunnel junction at low temperature to prevent junction aging effects as experimentally verified in our previous works \cite{Fevrier_2018}. We use infrared radiation  $\lambda =1550 \, \mathrm{nm}$ corresponding to  $V_0=0.825 \,\mathrm{V}$. We found that PAT in MIM junctions can not be explained by the theory of quantum detection as expected by eq.~(\ref{eq.Tucker}). Knowing that the Fermi-liquid theory \cite{Pine_Nozieres_1966} gives lifetime of an excited electron of the order of $\tau \sim \tau_0 (\epsilon_{\mathrm{F}}/\delta \epsilon)^2$, the lifetime of quasiparticles excited by optical photons is more than eight order of magnitude shorter than the ones excited by microwaves. For electrons with energy $\delta \epsilon = eV_0 =0.825 \, \mathrm{eV}$ above the Fermi energy, we estimate $\tau \sim 100 \, \mathrm{fs}$. This lifetime can only be accessed by  time-resolved photoemission spectroscopy \cite{Fann_1992,Campillo_1999}. As a consequence, the tunnel spectroscopy under continuous wave illumination can hardly reveal non-thermal distributions \cite{Sivan2019}. As a result, our data are well described by thermal distribution functions characterized by a temperature in each electrode and  highlight a tunneling Seebeck effect. \\


\begin{figure}[htbp!]
\begin{center}
\includegraphics[width=7cm]{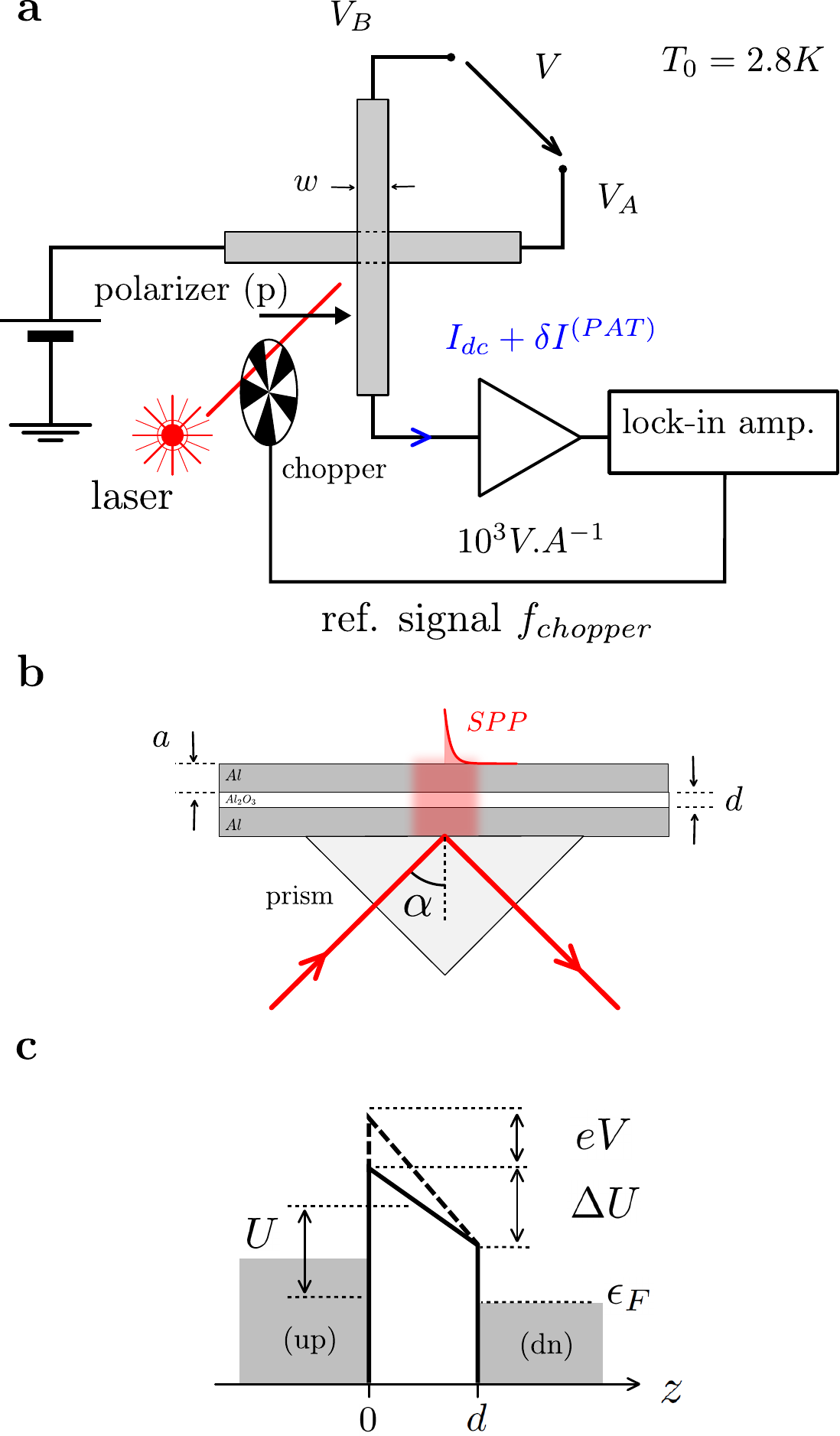}
\end{center}
\caption{\textbf{a} Experimental setup used to measure the photon assisted tunneling induced by a laser illumination at infrared frequencies. \textbf{b} We consider a Kretschmann configuration where a right angle prism allowed to enhance the coupling between the tunneling current and the light via the excitation of surface plasmons. \textbf{c} The tunnel barrier is modeled by a trapezoidal barrier. \label{fig:setup}}
\end{figure}

\noindent The article is organized as follows. The sample and the experimental setup  are described in section II. To increase the amount of radiation absorbed by the MIM, we convert the far field radiation into a local electromagnetic mode by means of a surface plasmon polariton (SPP) resonance. We probe the SPP resonance by PAT in section III. We then measure the right and left hand sides of eq.~(\ref{eq.Tucker}) independently. We discuss the validity of eq.~(\ref{eq.Tucker}) in the optical regime by using the Simmons tunnel barrier model in section IV \cite{Simmons_1964}.  In order to probe in more detail the interplay between illumination and the electron distribution in metallic thin films, we compare our measurement with the two possible senari: the hot carrier distribution functions characterized by a step like function at optical energies and the equilibrium distribution function at different temperatures.

\section{Sample fabrication and experimental setup}  
\begin{figure}[htbp!]
\begin{center}
\includegraphics[width=7cm]{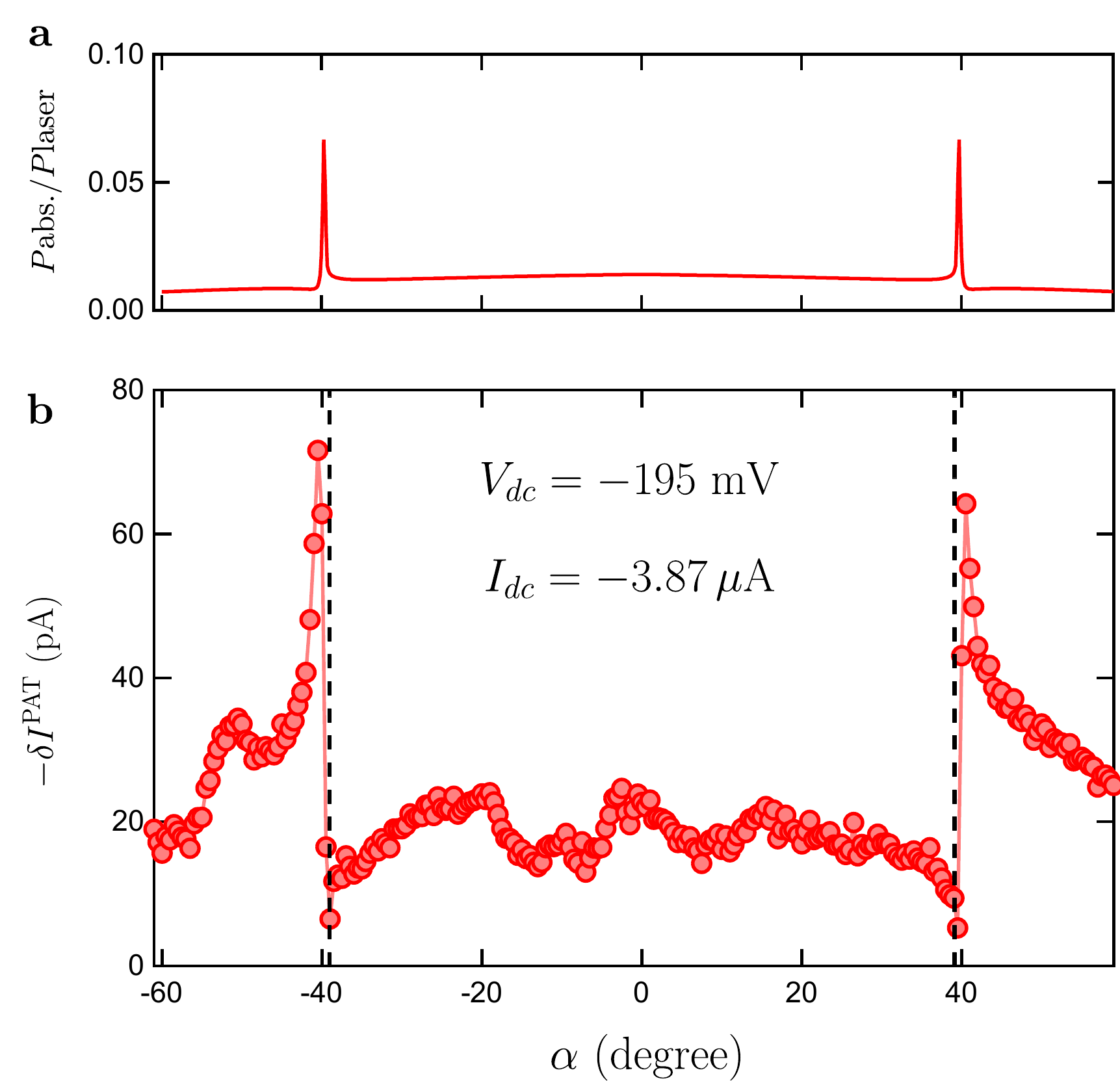}
\end{center}
\caption{\textbf{a}. Theoretical value of the ratio between the absorbed power $P_{\mathrm{abs.}}$ and the laser power $P_{\mathrm{laser}}$ as a function of the angle $\alpha$. \textbf{b} Detection of the SPP resonance by measuring the PAT at room temperature.\label{fig:PAT_angle_300K}}
\end{figure}

The sample is a $100 \times 100 \, \mu \mathrm{m}^2$ metallic cross-junction deposited on a standard microscope cover glass. The aluminum electrodes thickness is $a=8 \, \mathrm{nm}$ and the tunneling barrier is obtained by Al oxidation performed in a glow discharge. The $I(V)$ characteristics fit \cite{Simmons_1964} enables to determine the barrier characteristics depicted in Fig.~ \ref{fig:setup} \textbf{c}: thickness $d=1.6 \, \mathrm{nm}$, barrier height $U_0=2.9 \, \mathrm{eV}$ and barrier asymmetry $\Delta U_0=-0.78 \, \mathrm{eV}$.  A right angle prism is glued with a UV-curing glue on the rear of the glass to allow the SPP excitation by the incoming optical radiation in a Kretschmann configuration \cite{Kretschmann_1972} (see Fig.~ \ref{fig:setup} \textbf{b}). The experimental setup is shown in Fig.~ \ref{fig:setup}\textbf{a}. The junction is mounted on a rotating stage in a cryogen-free refrigerator at $T_0=2.8 \, \mathrm{K}$ enabling the excitation of the SPP resonance (SPR). The characteristics of the junction is measured using a standard four-point probe method simultaneously with the photon assisted current. The current generated due to the radiation is given by a lock-in amplifier referenced to the optical chopper at frequency $f_{\mathrm{chopper}}$ ranging from $80 \, \mathrm{Hz}$ to $900 \, \mathrm{Hz}$. The SPP mode is transverse magnetic. It is excited by a p-polarized light at the resonance angle $\arcsin (1/n) \simeq 41.1 ^\circ$ with $n=1.52$ the refractive index of the cover glass (see Appendix A). The p-polarization is set by adjusting a linear polarizer in the fridge at room temperature and small bias voltage before cooling. Figure \ref{fig:PAT_angle_300K} clearly shows the effect of the field enhancement due to surface plasmon resonance on the PAT \cite{Berthold_1985} at $\alpha_{res}= \pm 39.7 ^\circ$. In the following, we will consider PAT current induced by this resonance. It corresponds to an enhancement of the electromagnetic field in the tunnel junction by a factor 3.

\section{Photon-assisted tunneling at infrared frequencies}

The responsivity $\mathcal{R}$ of a rectifier is given by the ratio between the photon-assisted current $\delta I$ and the optical absorbed power $P$. Let's assume that optical excitation induces an ac signal of amplitude $V_{\mathrm{ac}}$  at frequency $\omega$. The related ac current is given by $I_{\mathrm{ac}}=G(\omega)V_{\mathrm{ac}}$ where $G(\omega)$ stands for the conductance at frequency $\omega$. In the classical limit of small ac excitation ($\hbar \omega \ll eV_{\mathbf{dc}}$ and $V_{\mathrm{ac}} \ll V_{\mathrm{dc}}$), the conductance is associated to the adiabatic response of  the $I(V)$ characteristics $G(\omega)=I'(V)=\frac{dI}{dV}$ and the time-averaged current is proportional to its second derivative $\delta I=I''(V_{\mathrm{dc}}) \times V_{\mathrm{ac}}^2 /2$. Surprisingly this classical limit seems to account for the experimental results of Ward \textit{et al} \cite{Ward_2010} out of its regime of validity $V_0 \gg  V_{\mathrm{dc}}$.  With the hypothesis that the optical absorbed power is equal to the electrical one $P=G(\omega) V_{\mathrm{ac}}^2$, the responsivity can be rewritten by $\mathcal{R}=\frac{I''(V_{\mathrm{dc}})}{2I'(V_{\mathrm{dc}})}$ as for a classical diode. Following the same assumption and using the Tucker's description of the electronic transport in a metallic tunnel junction at finite frequencies, the conductance is given by $G(\omega)=\frac{I(V_{\mathrm{dc}}+V_0)-I(V_{dc}-V_0)}{2V_0}$ and the responsivity in the quantum regime reads \cite{Grover2013}:

\small
\begin{equation}\label{eq.responsively}
\mathcal{R}(V_{\mathrm{dc}})=\frac{1}{V_0} \frac{I_0(V_{\mathrm{dc}}+V_0)+I_0(V_{dc}-V_0)-2I_0(V_{\mathrm{dc}})}{I_0(V_{\mathrm{dc}}+V_0)-I_0(V_{\mathrm{dc}}-V_0)}.
\end{equation}

\begin{figure}[htbp!]
\begin{center}
\includegraphics[width=7cm]{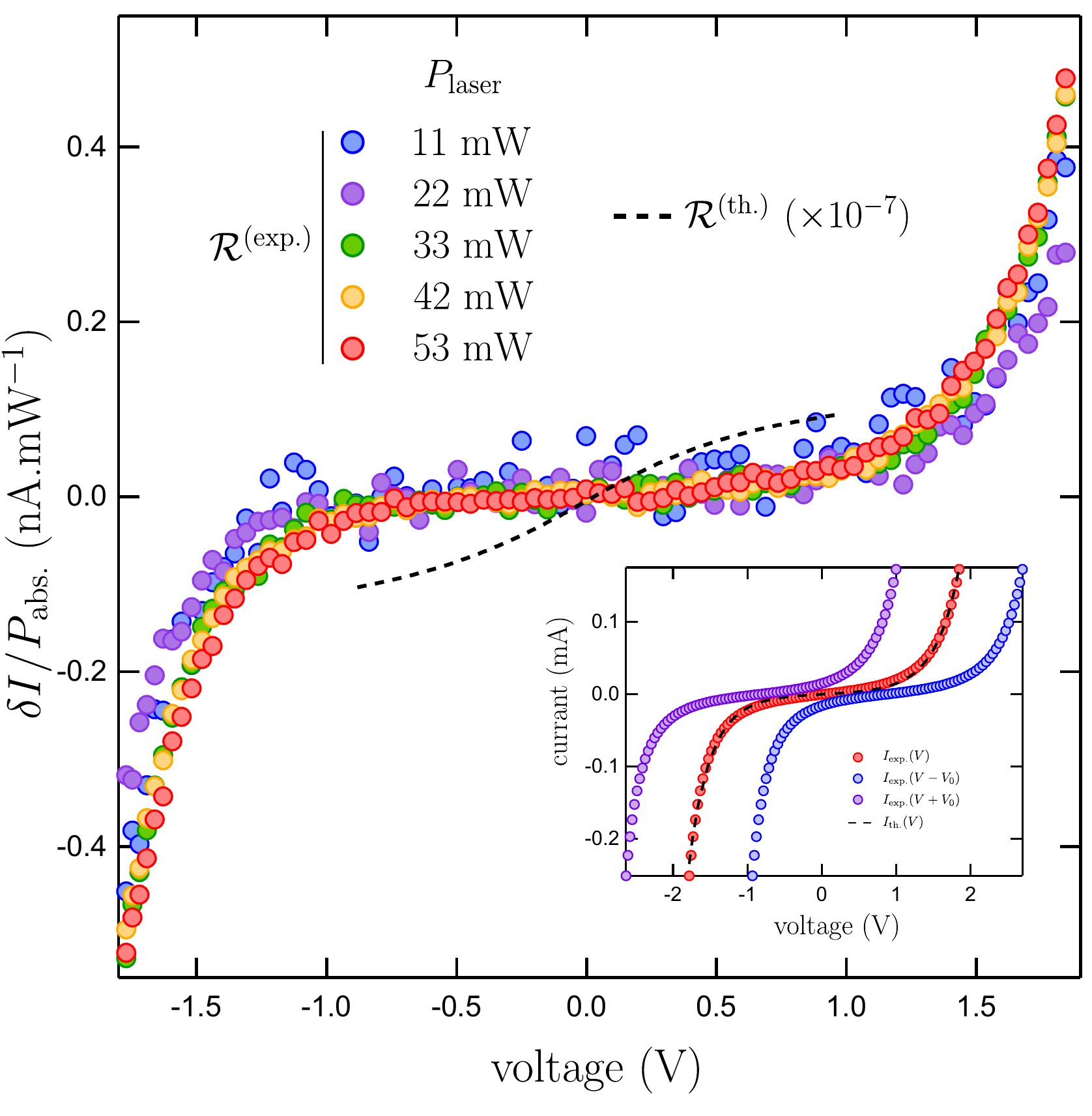}
\end{center}
\caption{Normalized PAT current $\delta I^{\mathrm{(PAT)}}/P_{\mathrm{abs}}$  for laser power ranging between $11 \, \mathrm{mW}$ and $53 \, \mathrm{mW}$. The width of the laser spot on the sample is $w_{\mathrm{laser}}=650 \, \mu \mathrm{m}$. The measurement is performed at $T_0=2.8 \, \mathrm{K}$ at the SPP resonance $\theta_{\mathrm{res}}=-39^{\circ}$. \textit{Inset bottom right:} $I(V)$ and $I(V\pm V_0)$ characteristics of the tunnel junction measured at low temperature $T_0=2.8 \, \mathrm{K}$. The fit of $I(V)$ with the Simmons theory gives $d=1.6 \, \mathrm{nm}$, $U_0=2.9 \, \mathrm{eV}$, $\Delta U_0=-0.78 \, \mathrm{eV}$.\label{fig:PATI_volt_pwr}}
\end{figure}

\noindent
Figure \ref{fig:PATI_volt_pwr} shows the experimental  responsivity $\mathcal{R}^{(\mathrm{exp.})}(V_{dc})=\delta I^{\mathrm{(PAT)}}/P_{\mathrm{abs.}}$ measured as a function of the bias voltage for different laser powers at SPR. The absorbed power at SPR is proportional to the laser power: $P_{\mathrm{abs.}}\simeq 0.07 P_{\mathrm{laser}}$ (see Fig.~ \ref{fig:PAT_angle_300K}\textbf{a} and Appendix A). As shown in eq.~(\ref{eq.responsively}), the responsivity does not depend on the illumination power. The conversion efficiency of the metallic tunnel junction is independent of the incident optical power and acts as linear photon detector for the investigated power range ($P_{\mathrm{laser}}< 100 \, \mathrm{mW}$). We estimate the maximal experimental responsivity at SPR: $\mathcal{R}^{(\mathrm{exp.})}(V_{\mathrm{dc}}=1.7 \, \mathrm{V}) \sim 400 \mathrm{nA}.\mathrm{W}^{-1} $. By using the measured $I(V)$ characteristics between $-1.7 \, \mathrm{V}$ and $1.8 \,\mathrm{V}$ (see inset Fig.~ \ref{fig:PATI_volt_pwr}), we estimate the theoretical responsivity $\mathcal{R}(V_{\mathrm{dc}})$ between $-0.88 \, \mathrm{V}$ and $0.97 \, \mathrm{V}$. We see in Fig.~ \ref{fig:PATI_volt_pwr} that $\mathcal{R}^{(\mathrm{exp.})}$ does not have the same voltage dependence (black dashed line) and is more than seven orders of magnitude smaller than the theoretical value given by eq.~(\ref{eq.responsively}). This requires some more investigation which we do here below.

\section{Validity of the Tucker formula}

Equation (\ref{eq.responsively}) is  convenient because it directly gives the photon responsivity of the junction from its electrical characteristics $I(V)$.  However, this description is based on two strong hypothesis: (i) the photon-assisted processes can be described by an optical voltage biasing of the junction and (ii) the absorbed power is given by the electric resistance of the junction evaluated at optical frequency. Let us briefly show how to establish eq.~ (\ref{eq.Tucker}). The effect of voltage bias $V(t) = V_{\mathrm{dc}} + V_{\mathrm{ac}} \cos \left( \frac{eV_0}{\hbar} t \right)$ due to the dc bias and the laser illumination is to shift adiabatically the Fermi sea of the upper reservoir sustaining the surface plasmon while the lower reservoir is grounded. The resulting time averaged out-of-equilibrium distribution function $\tilde{f}$ in the upper reservoir is then given by the Tien-Gordon formula \cite{Tien_1963,Blanter_2000}:

\begin{equation}\label{eq.distrib}
\tilde{f}(\epsilon,T_0)=\sum_{n=-\infty}^{+\infty} \left| J_n \left(\frac{V_{\mathrm{ac}}}{V_0} \right) \right|^2 f(\epsilon+neV_0,T_0),
\end{equation}

\noindent where  $J_n$ is the $n^{\mathrm{th}}$  Bessel's function of the first kind and $f(\epsilon,T_0)=1/\left(1+\exp((\epsilon-\epsilon_F)/k_BT_0)\right)$ is the Fermi–Dirac distribution with $\epsilon_F$ the Fermi energy and $T_0$ the electron temperature. Using a Landauer-B\"{u}ttiker  scattering approach, we express the time-average PAT current as:

\begin{equation}\label{eq.LB1}
\begin{split}
\delta I(V_{\mathrm{dc}},V_{\mathrm{ac}})&=\frac{e}{h} \int d\epsilon \mathcal{T}(\epsilon,eV_{\mathrm{dc}}) \left[\tilde{f}(\epsilon -eV_{\mathrm{dc}},T_0) \right.\\
& \left.- f(\epsilon -eV_{\mathrm{dc}},T_0)\right].
\end{split}
\end{equation}

\noindent Using $\sum_{n} \left| J_n \right|^2=1$ and assuming that transmission is voltage independent $\mathcal{T}(\epsilon,eV_{\mathrm{dc}})=\mathrm{cte}$ , one deduces:

\begin{equation}\label{eq.LB2}
\delta I=\sum_{n=-\infty}^{+\infty} \left| J_n \left(\frac{V_{\mathrm{ac}}}{V_0} \right) \right|^2 I\left( V_{\mathrm{dc}}+nV_0 \right) - I\left( V_{\mathrm{dc}} \right),
\end{equation}

\begin{figure}[htbp!]
\begin{center}
\includegraphics[width=7cm]{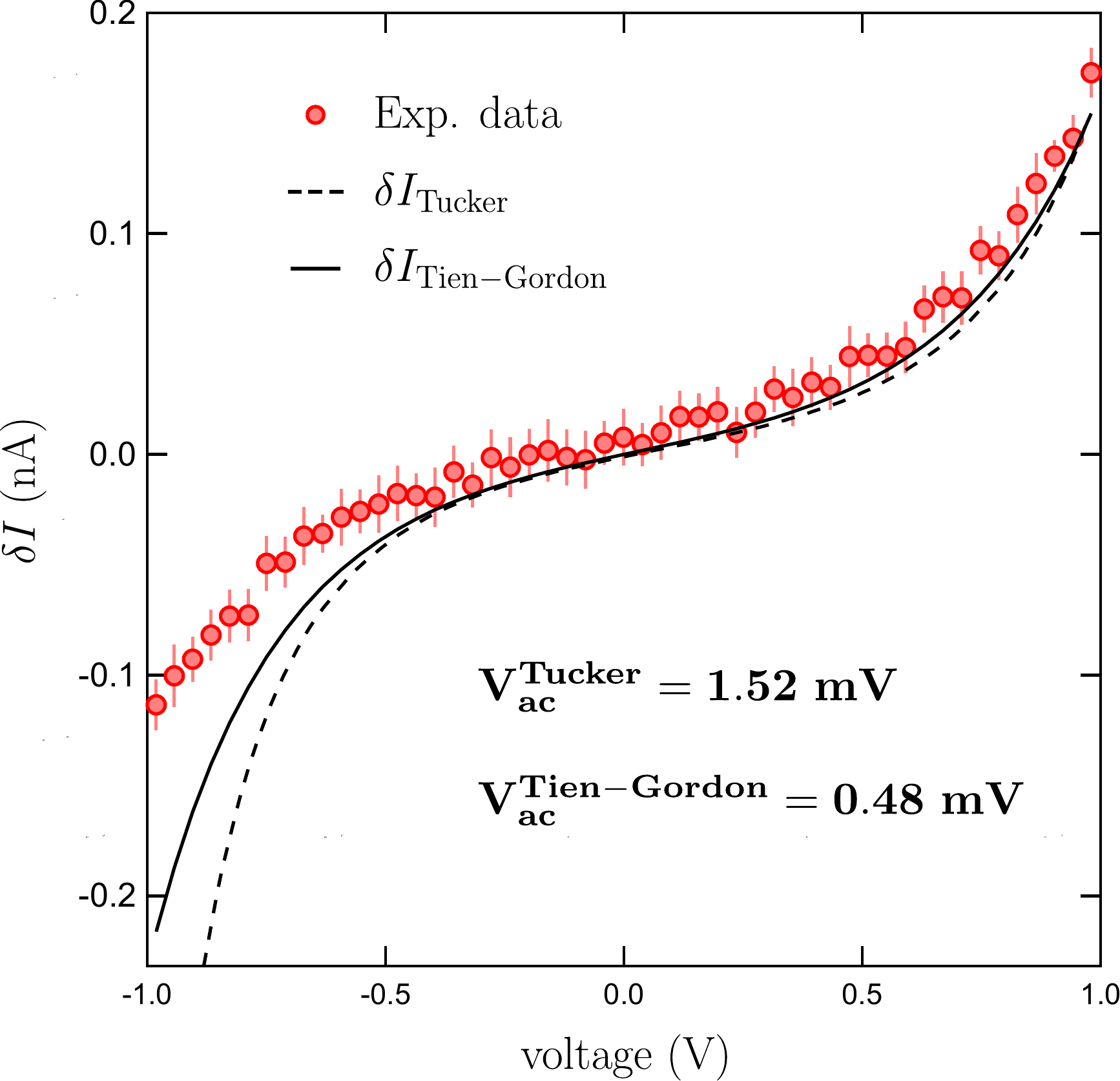}
\end{center}
\caption{PAT current as a function of dc voltage bias under an illumination at $\lambda=1550 \, \mu \mathrm{m}$ and $T_0=2.8 \, \mathrm{K}$. The light intensity is $16 \, \mathrm{W}.\mathrm{cm}^{-2}$. The black dashed line corresponds to the measured right hand side of eq.~(\ref{eq.Tucker}). The only fitting parameter is the value of $V_{\mathrm{ac}}=V_{\mathrm{op}}^{\mathrm{Tucker}}$ which is adjusted with the slope at zero bias. The black solid line corresponds to the numerical simulation of eq.~(\ref{eq.LB1}) using the barrier parameters $U \simeq 2.9 \, \mathrm{eV}$, $\Delta U \simeq -0.78\, \mathrm{eV}$ and  $d \simeq 1.6 \, \mathrm{nm}$.   \label{fig:PAT_Tucker}}
\end{figure}

\noindent If one only considers the one photon apsorption/emission processes at energies $e \left(V_{dc} \pm V_0 \right)$ ($V_{\mathrm{ac}} \ll V_0$), eq.~(\ref{eq.LB2}) corresponds to eq.~(\ref{eq.Tucker}) by retaining the lowest order terms in $V_{\mathrm{ac}}/V_0 $ . Note that eq.~(\ref{eq.LB1}) is explicitly not gauge invariant \cite{Aguado2003}. This hypothesis is valid in the microwave regime where the alternative electric field is well described by a voltage drop across the junction (one electrode is effectively grounded). However, it remains a strong assumption in optics where the lower electrode is not grounded at optical frequencies and the associated vector potential does play a role. In what follows, we  focus on the experimental verification of eq.~ (\ref{eq.Tucker}). To do so, we probe $\delta I$ in a range of $\pm V_{\mathrm{exp}}$ and  the $I(V)$ characteristics in a range of $\pm \left(V_{\mathrm{exp}}+V_0 \right) \simeq \pm 1.8 \, \mathrm{V}$. Figure \ref{fig:PAT_Tucker} shows the PAT current measured by optically chopping of the laser beam at $\sim 120 \, \mathrm{Hz}$. The data are compared to the TF photon assisted tunneling current $\delta I_{\mathrm{Tucker}}$ given by eq.~ (\ref{eq.Tucker}). Note that this quantity is directly measured. The only free parameter is the optical voltage $V_{\mathrm{ac}}$ which is adjusted with the slope of $\delta I^{\mathrm{(PAT)}}$ at zero dc bias. We then deduce an optical voltage $V_{\mathrm{ac}}^{(\mathrm{Tucker})}=1.52 \, \mathrm{mV}$. This value is overestimated by more than  four orders of magnitude in view of the laser power (according to Appendix A $P_{\mathrm{laser}}=53 \, \mathrm{mW}$ corresponds to  $V_{ac}\simeq 70 \, \mathrm{nV}$). More importantly equation (\ref{eq.Tucker}) does not account for the experimental data neither qualitatively or quantitatively. As mentioned above, this equation assumes that tunnel transmission is voltage independent: $\mathcal{T}(\epsilon,eV_{\mathrm{dc}})=\mathcal{T}_0(\epsilon)$. This is not the case at bias voltage of the order of the tunnel barrier height. Instead, the transmission $\mathcal{T}(\epsilon,eV_{\mathrm{dc}})$ is calculated by modelling the junction by a trapezoidal barrier \cite{Simmons_1964} (see inset Fig.~\ref{fig:PATI_volt_pwr} and Appendix B). A numerical evaluation of eq.~(\ref{eq.LB1}) gives a smaller but still unrealistic optical voltage $V_{\mathrm{ac}}^{(\mathrm{Tien-Gordon})}=0.48 \, \mathrm{mV}$. In conclusion, the fitting parameters using TF are not physical. Our experiment confirms that the  description of the photon-assisted tunneling as an optical rectification cannot explain the photon-assisted processes in a tunnel junction. A more realistic description is to consider the Boltzmann equation (BE) describing electron dynamics in metallic systems \cite{Voisin_2000}.

\section{Hot-carrier vs thermal distribution}

\begin{figure}[htbp!]
\begin{center}
\includegraphics[width=7cm]{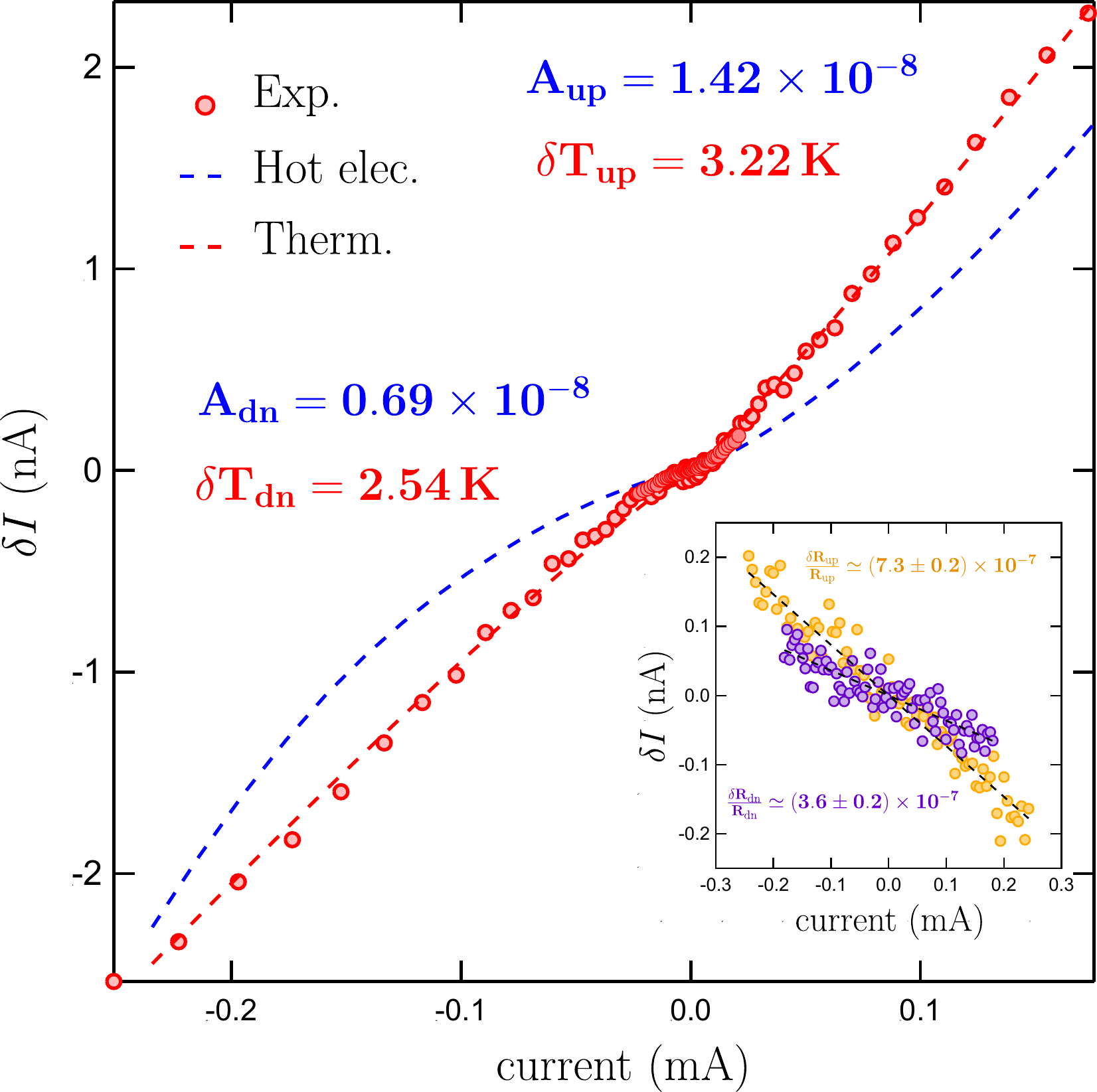}
\end{center}
\caption{PAT current as a function of dc current with the same experimental settings as Fig.~\ref{fig:PAT_Tucker}. Symbols are experimental data. Blue dashed line is the numerical simulation from eqs.~ (\ref{eq.distrib_ph_abs}) and (\ref{eq.LB3}) dealing with hot carrier distribution in the electrodes whereas the red dashed line corresponds to two different thermal distributions in the contacts described. \textit{Inset}: photon-induced current measured in the upper (orange circles) and lower contacts (purple circles).\label{fig:PATI_current_fit}}
\end{figure}

The PAT current of eq.~(\ref{eq.LB1}) can be re-written in terms of the deviation of the electron distribution from the equilibrium:

\begin{equation}\label{eq.LB3}
\begin{split}
\delta I(V_{\mathrm{dc}},\delta f_{\mathrm{up}},\delta f_{\mathrm{dn}})=&\frac{e}{h} \int d\epsilon \; \mathcal{T}(\epsilon,eV_{\mathrm{dc}}) \times\\
 &\left[\delta f_{\mathrm{up}}(\epsilon +eV_{\mathrm{dc}}) - \delta f_{\mathrm{dn}}(\epsilon)\right],
\end{split}
\end{equation}

\noindent where $\delta f_{\mathrm{up/dn}}=f(\epsilon)-f(\epsilon,T_0)$ states for the deviation of the electron distribution from the equilibrium distribution at base temperature $T_0$. This steady-state non-equilibrium electron distribution function under continuous wave illumination is  given by the master equation:

\begin{equation}\label{eq.BE1}
\left(\frac{\partial f}{\partial \epsilon}\right)_{\mathrm{e-ph}}+\left(\frac{\partial f}{\partial \epsilon}\right)_{\mathrm{e-e}}+\left(\frac{\partial f}{\partial \epsilon}\right)_{\mathrm{exc}}=0,
\end{equation}

\noindent where $f$ is the electron distribution function, the first term stands for electron-phonon energy transfer between electrons and lattice, the second stands for electron-electron thermalization (characterized by an electron collision time of the order of few 100 fs)  while the third term describes the excitation due to photon absorption in the contacts. The first two terms will mainly lead to an equilibrium distribution characterized by an effective temperature. The excitation term could however lead to the existence of a small fraction of hot electron. we evaluate below what would be the effect of a hot electron distribution on the PAT. It can be evaluated by considering a  detailed balanced equation and the Fermi golden rule (Appendix C):

\begin{equation}\label{eq.distrib_ph_abs}
\begin{split}
\delta f_{\mathrm{exc}}=\left(\frac{\partial f}{\partial \epsilon}\right)_{\mathrm{exc}}=&A_{\mathrm{exc}} \left[ \sqrt{\epsilon-eV_0} f(\epsilon-eV_0) \left(1-f(\epsilon) \right) \right.\\
& \left.-\sqrt{\epsilon+eV_0} f(\epsilon) \left(1-f(\epsilon+eV_0) \right) \right],
\end{split}
\end{equation}

\noindent where $A_{\mathrm{exc}}$ is a constant proportional to the energy absorbed by electrons. In the relaxation time approximation for the electron-electron thermalization , we deduce the time-average PAT current of eq.~(\ref{eq.LB1}) due to hot-carrier contribution by considering the deviation of the electron distribution  $\delta f_{\mathrm{up/dn}}=\delta f(\epsilon,A_\mathrm{up/dn}) \propto \left(\frac{\partial f}{\partial \epsilon}\right)_{\mathrm{up/dn}}$. $A_{\mathrm{up}/\mathrm{dn}}$ is a constant proportional to the energy absorbed in the upper/lower electrode. It is convenient to characterize this amount of energy by expressing it in terms of equivalent temperature of excitation $\delta T_{\mathrm{exc}}$. To do so, we consider  a Fermi-Dirac distribution at $T_0+\delta T_{\mathrm{exc}}$ containing the same energy as the non-equilibrium distribution function:

\begin{equation}\label{eq.Texc}
\int \epsilon \rho(\epsilon) \left[ f(\epsilon,T_0)+\delta f_{\mathrm{exc}} \right]\, d\epsilon=\int \epsilon \rho(\epsilon) f(\epsilon,T_0+ \delta T_{\mathrm{exc}})\,  d\epsilon,
\end{equation}

\noindent where $\rho(\epsilon) \propto \sqrt{\epsilon}$ is the density of states of the free electron gas. Figure \ref{fig:PATI_current_fit} shows the photon-assisted current $\delta I^{(ph)}$  as a function of dc current flowing through the junction biased between $-1.78 \, \mathrm{V}$ and $+1.84 \, \mathrm{V}$. It is mainly proportional to the current. Dashed blue line in figure \ref{fig:PATI_current_fit} corresponds to the theoretical prediction given by eqs.~(\ref{eq.distrib_ph_abs},\ref{eq.LB3}). We clearly see that the proportionality cannot be explained by a hot-carrier distribution function. We fit $\delta I(I_{\mathrm{dc}})$ by adjusting the absorption coefficients  $A_{\mathrm{up}} \sim 1.42 \times 10^{-8}$ and $A_{\mathrm{dn}} \sim 0.69 \times 10^{-8}$  with an optimization software. Note that $A_{\mathrm{up}}>A_{\mathrm{dn}}$ is in agreement with the fact that upper electrode experiences the field enhancement of the surface plasmon. They corresponds to equivalent temperatures of excitation of $\delta T_{\mathrm{exc,up}}=0.30 \, \mathrm{K}$ and $\delta T_{\mathrm{exc,dn}}= 0.15 \, \mathrm{K}$  respectively. Even if this term leads to a Tien-Gordon-like distribution  given by eq.~(\ref{eq.distrib}), the hot carrier distribution remains extremely small because of electron heating and  heat leakage to the lattice as studied by Dubi and Sivan \cite{Sivan2019,Bouhelier_2021}. Moreover, the equivalent temperature of excitation is smaller than $0.3 \, \mathrm{K}$ and we see in the following that this temperature remains smaller than the temperature induced by the power absorption in the lead. It follows that the photon absorption mainly induces heating rather than generating hot carriers. In the following, we  analyze the PAT current as resulting from two different thermal distributions in the contacts  and can be refereed as a the tunneling Seebeck effect \cite{Stovneng_1990}. However, unlike the Seebeck effect which depends on the metallic density of states of the electrode materials, the tunneling Seebeck effect depends on the energy dependence of the tunneling transmission and so that on the non linearity of the $I(V)$ characteristic \cite{Smith_1980}. The PAT current is then described by eq.~(\ref{eq.LB3}) with the deviation of the electron distribution $\delta f_{\mathrm{up/dn}}=\delta f(\epsilon,T_\mathrm{up/dn})=f(\epsilon,T_0+\delta T_{\mathrm{up/dn}})-f(\epsilon,T_0)$. $T_{\mathrm{up}/\mathrm{dn}}$ is the temperature of upper/lower electrode.

 Figure \ref{fig:PATI_current_fit} shows the proportionality between the photon assisted current and the current. This proportionality is explained by a differential heating between the upper and lower electrodes. Our data are in good agreement with the heating description (red dashed line in Fig.~\ref{fig:PATI_current_fit}). We have checked that the PAT measured across the junction is one order of magnitude greater than the photon assisted current in each electrode.  The cross geometry of the tunnel junction allows us to measure the photon-assisted current in the upper/lower electrode by using a current polarization of the upper/lower electrode. The power supply is connected on the $V_B$  contact (\textit{resp.} $V_A$) to measure the upper (\textit{resp.} lower) electrode (see Fig.~\ref{fig:PAT_angle_300K}) while the current amplifier is connected to the opposite side of the electrode. Inset of figure  \ref{fig:PATI_current_fit} shows the photo-assisted current measured  in the two electrodes independently. This photon assisted current is attributed to a variation of the aluminum conductivity. The Drude conductivity is indeed proportional to the relaxation time which depend on the temperature so that a laser induced temperature modulation will induce a conductance modulation.  The energy and voltage dependent transmission $\mathcal{T}(\epsilon,eV_{\mathrm{dc}})$ of the barrier is deduced from the fit of the $I(V)$ curve. The fit of the curve $\delta I^{(ph)}(I)$ gives $\delta T_{\mathrm{up}} \simeq 3.22\, \mathrm{K} $ and $\delta T_{\mathrm{dn}} \simeq 2.54\, \mathrm{K}$. The temperature rise can be associated with the optical power absorbed by the electrodes. The junction in the Kretschmann configuration forms a well defined plasmonic multilayer stack  and the absorbed power at the resonance angle $\alpha_{res}=-39.7 ^{\circ}$ are evaluated to $P_{\mathrm{up}}\simeq 265 \, \mathrm{\mu W}$ and $P_{\mathrm{dn}}\simeq 84 \, \mathrm{\mu W}$ respectively (Appendix A). We neglected the power leakage due to thermal conduction of the contacts and identify the absorbed power and the power dissipated to the lattice of volume $\Omega=a \times w \times w_{\mathrm{laser}} \simeq 5.2 \times 10^{-16} \, \mathrm{m}^{-3}$. The temperature rise in the electrodes is thus estimated to $\delta T_{\mathrm{up/dn}}=\left( \frac{P_{\mathrm{up/dn}}}{\Sigma \Omega}\right)^{1/5}$ where the electron-phonon coupling constant in aluminum \cite{Meschke_2004} is $\Sigma \simeq 0.3 \, \mathrm{nW}.\mu\mathrm{m}^{-3}.\mathrm{K}^{-5}$. It gives $\delta T_{\mathrm{up}} \sim 1.7 \, \mathrm{K}$ and $\delta T_{\mathrm{dn}} \sim 1.0 \, \mathrm{K}$. Note that the estimated temperature rise is slightly underestimated. It can be attributed to the value of the dielectric constant that we have considered \cite{Rakic_1995} knowing we consider here an unusual ultra-thin aluminum film. The photon-assisted tunneling is strictly speaking a thermo-assisted tunneling. Our measurement could also be sensitive to the variation of the electrode conductance due to their temperature modulation. Thanks to the cross-geometry of the junction, we have checked that this effect is negligible compare to the PAT measured on the junction. The inset of  Fig.~\ref{fig:PATI_current_fit} shows that the photon-induced current in the electrodes is lower by more than one order of magnitude compared to the PAT current. The modulation of the light power by the chopper (laser on/off) induces a modulation of the resistance of the electrodes: $R_{\mathrm{up/dn}}^{\mathrm{on}}=R_{\mathrm{up/dn}}^{\mathrm{off}}+ \delta R_{\mathrm{up/dn}}$. When the electrode is voltage bias, this translates in a  photon-induced current  given by $\delta I = I^{\mathrm{on}} -I^{\mathrm{off}}=-\frac{\delta R_{\mathrm{up/dn}}}{R_{\mathrm{up/dn}}} I$. The inset of Fig.~\ref{fig:PATI_current_fit} shows that the photon induced current in the electrodes corresponds to a relative increase of their resistance ($\delta R>0$) \cite{Herzog_2014}: $\eta_{\mathrm{up}}=\frac{\delta R_{\mathrm{up}}}{R_{\mathrm{up}}}\simeq (7.3\pm 0.2) \times 10^{-7}$ and $\eta_{\mathrm{dn}}=\frac{\delta R_{\mathrm{dn}}}{R_{\mathrm{dn}}}\simeq (3.6\pm 0.2) \times 10^{-7}$ .  Using the Sommerfeld expansion in the electrical conductivity and assuming a linear dependence of the relaxation time, we expect a quadratic temperature dependence of the conductivity so that $\frac{\eta_{\mathrm{up}}}{\eta_{\mathrm{dn}}}=\left( \frac{\delta T_{\mathrm{up}}}{\delta T_{\mathrm{dn}}} \right)^2 \simeq 2.0 \pm 0.3$ (see Appendix D). We then deduce a ratio $\sqrt{\frac{\eta_{\mathrm{up}}}{\eta_{\mathrm{dn}}}} \simeq 1.4 \pm 0.2$ which is consistent with the measured ratio $\frac{\delta T_{\mathrm{up}}}{\delta T_{\mathrm{dn}}} \simeq 1.27$.

\section{Conclusion}

Optical rectification in metallic tunnel junctions produces dc photocurrent assuming that illumination acts as an ac voltage at optical frequencies across the nonlinear tunnelling conduction. The photon-assisted transport theory in tunnel junctions states that the dc photocurrent is a well known function of the $I(V)$ characteristics on a voltage range of the photon voltage $\hbar \omega/e$. This theoretical prediction is based on the creation of a non-thermal hot carrier distribution function in the metals under optical illumination. We have shown experimentally that this is incorrect in a planar tunnel junction. By comparing the photocurrent generated in a planar junction in the Kretschmann configuration and the measured $I(V)$ characteristics, we have demonstrated that the photocurrent can be described by heating and thermal distribution functions instead. While we have only considered a planar tunnel junction without nanoantenna, the question of the generation of hot electrons in metallic nanostructures for optical rectification or optical detection still raises theoretical questions. As pointed out by I\~{n}arrea et \textit{al.} in the case of THz detectors in double barrier systems  \cite{Inarrea_1994}, it is crucial to calculate the transmission coefficient by considering the coupling term $\mathbf{A}.\mathbf{p}$  where $\mathbf{A}$ is the vector potential of the electromagnetic field and $\mathbf{p}$ the electronic momentum operator. Note that the coupling between electrons and the electromagnetic field no longer takes place in a single region of the structure but interacts with the barrier. In conclusion, the planar tunnel junction is not a good candidate to highlight the hot carrier distribution in illuminated metallic thin films. The metallic density of states is indeed to smooth and it does not allow the realization of the necessary energy filter as it is in a molecular junction \cite{Dubi_2022}.

\section*{Appendix A: Plasmonic multilayer stack}

\begin{figure}[htbp!]
\begin{center}
\includegraphics[width=6cm]{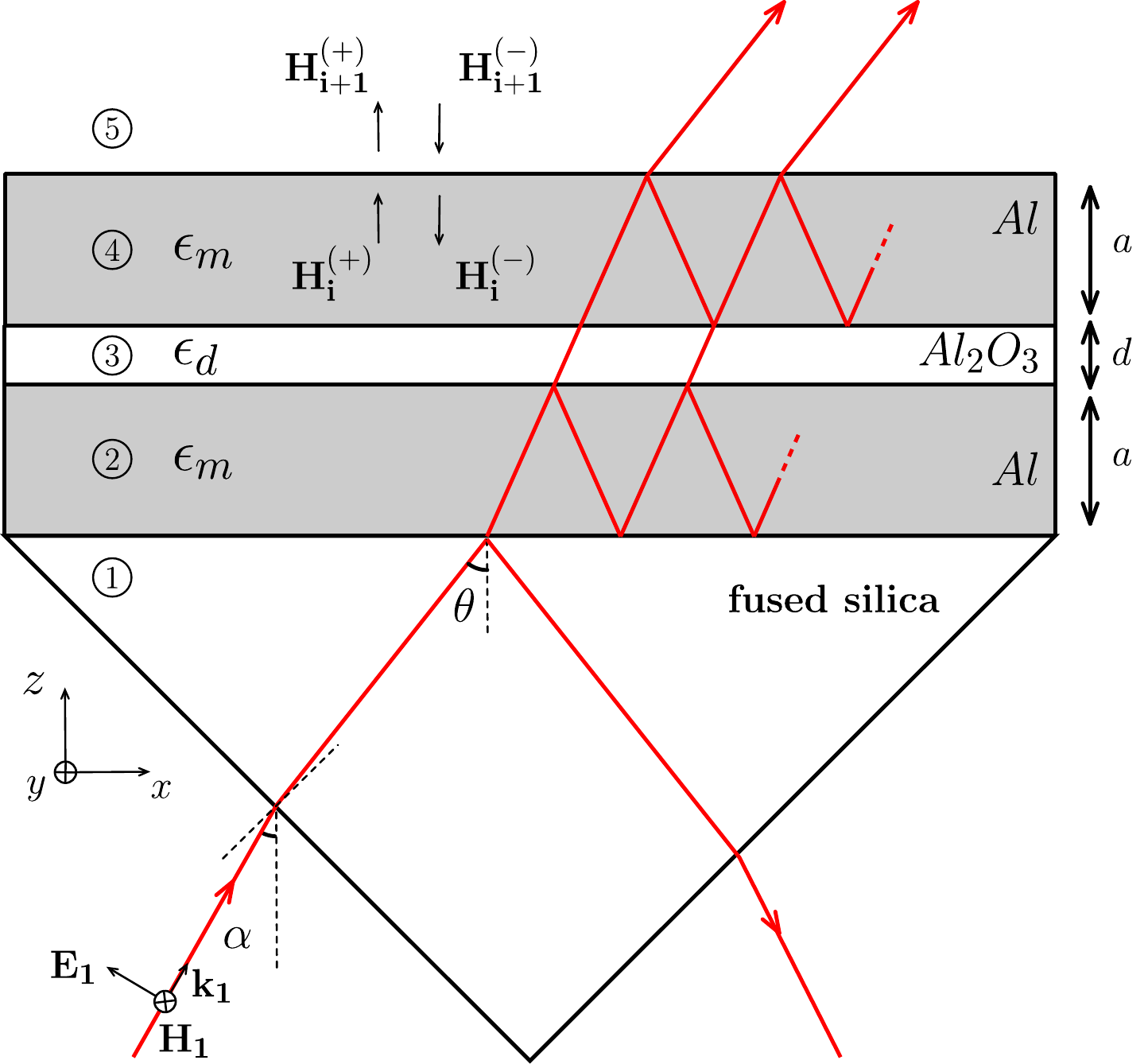}
\end{center}
\caption{Schematics of the MIM stack. $\alpha$ is the incident angle in the vacuum whereas $\theta$ is the angle in the prism. The Snell–Descartes laws give $\sin \left( \frac{\pi}{4}-\alpha \right)=n \sin \left( \frac{\pi}{4}-\theta \right)$. \label{fig:Kretchmann}}
\end{figure}

We consider an $\mathrm{Al/Al_2O_3/Al}$ MIM stack on top of a righ angle prism (see Fig.~\ref{fig:Kretchmann}). The thickness of the metallic films is $a=8\,\mathrm{nm}$ and that of dielectric barrier is $d \simeq 1.6\,\mathrm{nm}$. The  dielectric constants of metal and barrier are respectivelly $\epsilon_{m}=- 242+ 49.4 \mathrm{i}$ and $\epsilon_{m}=3.051+0.0629\mathrm{i}$ at $\lambda=1550 \mu \mathrm{m}$ \cite{Rakic_1995,Dodge_1986}.

\noindent We consider the Fresnel coefficients for a TM-polarised light (p polarized). For a single interface from (1) to (2) between two dielectric materials characterized by the dielectric constant $\epsilon_1$ and $\epsilon_2$ respectively, the reflexion and transmission Fresnel coefficients read:
\begin{equation}\label{eq.fresnel}
 r_{1 \hookleftarrow 2}= \frac{\epsilon_2k_{z1}-\epsilon_1k_{z2}}{\epsilon_2k_{z1}-\epsilon_1k_{z2}}, \; t_{1 \rightarrow 2}= \frac{2\epsilon_2k_{z1}}{\epsilon_2k_{z1}-\epsilon_1k_{z2}}.
\end{equation}

\noindent By using the Fresnel’s coefficients on the interfaces at $z=0,a,a+d,2a+d$, the amplitude of the incoming $H_{i}^{(-)}$ and outgoing  $H_{i}^{(+)}$ magnetic field are connected by a transfer matrix:

\begin{equation}\label{eq.transfer2}
 T_{i,i+1}=\left(
                         \begin{array}{cc}
                            t_{i \rightarrow i+1}- \frac{r_{i \hookleftarrow i+1} r_{i+1 \hookleftarrow i}}{t_{i+1 \rightarrow i}} & \frac{r_{i+1 \rightarrow i}}{t_{i+1 \rightarrow i}} \\
                           -\frac{r_{i \rightarrow i+1}}{t_{i+1 \rightarrow i}}& \frac{1}{t_{i+1 \rightarrow i}} \\
                         \end{array}
                       \right).
\end{equation}

\begin{figure}[htbp!]
\begin{center}
\includegraphics[width=7cm]{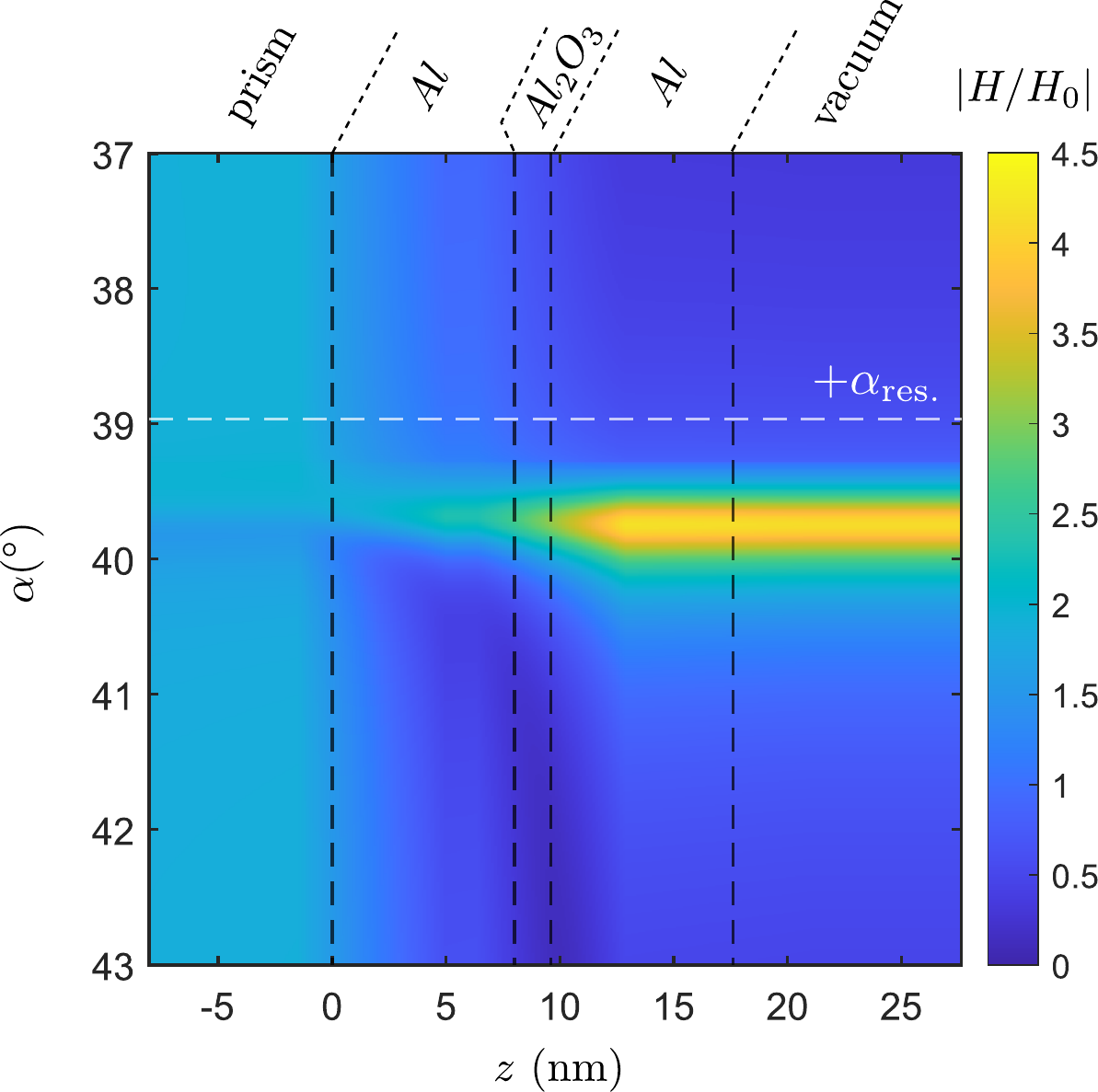}
\end{center}
\caption{Magnitude of the magnetic field inside the junction for different incident angle $\alpha$ at SPR. \label{fig:Kretchmann_Hfield}}
\end{figure}

\noindent The p-polarized magnetic fields $\mathbf{H}=\left( H^{(+)}_i  e^{ik_{zi} z } +H^{(-)}_i  e^{ -ik_{zi} z} \right) \mathbf{e_{\mathrm{y}}}$ in the different regions are defined by:

\begin{subequations}
\begin{align*}
\left(\begin{array}{c}
            H_{2}^{(+)} \\
            H_{2}^{(-)} \\
          \end{array}
        \right)&=T_{1,2} \times \left(
                    \begin{array}{c}
                      H_{1} \\
                      H_{1}^{(-)} \\
                    \end{array}\right),\\
\left(\begin{array}{c}
            e^{ ik_{z3} a} H_{3}^{(+)} \\
            e^{ -ik_{z3} a} H_{3}^{(-)} \\
          \end{array}
        \right)&=T_{2,3} \times  \left(
                    \begin{array}{c}
                      e^{ik_{z2} a } H_{2}^{(+)} \\
                      e^{-ik_{z2} a}H_{2}^{(-)} \\
                    \end{array}\right), \\
\left(\begin{array}{c}
            e^{ ik_{z4} (a+d)} H_{4}^{(+)} \\
            e^{-ik_{z4} (a+d)} H_{4}^{(-)} \\
          \end{array}
        \right)&=T_{3,4} \times  \left(
                    \begin{array}{c}
                      e^{ik_{z3} (a+d)} H_{3}^{(+)} \\
                      e^{-ik_{z3} (a+d)}H_{3}^{(-)} \\
                    \end{array}\right) , \\
\left(\begin{array}{c}
            e^{ik_{z5} (2a+d)} H_{5}^{(+)} \\
            0 \\
          \end{array}
        \right)&=T_{4,5} \times  \left(
                    \begin{array}{c}
                      e^{ik_{z4} (2a+d)} H_{4}^{(+)} \\
                      e^{-ik_{z4} (2a+d)} H_{4}^{(-)} \\
                    \end{array}\right), \\
\nonumber
\end{align*}
\end{subequations}
\noindent with,

\begin{subequations}
\begin{align*}
k_{\mathrm{x}}&=\sqrt{\epsilon_0} \left( \frac{\omega}{c} \right) \sin \theta,\\
 k_{zi}&=\sqrt{\epsilon_i \left( \frac{\omega}{c} \right)^2-k_{\mathrm{x}}^2} \hspace*{0.3cm} \forall i \in \left\{ 1,...,5  \right\}.\\
\nonumber
\end{align*}
\end{subequations}

\noindent The color plot Fig.~\ref{fig:Kretchmann_Hfield} shows the surface plasmon resonance (SPR) at the angle $\alpha_0=\pm 39.7 ^{\circ}$. This angle corresponds approximatively to the critical angle $\alpha_{\mathrm{c}}= \frac{\pi}{4}- \arcsin \left[ n \sin \left( \frac{\pi}{4} -\theta_{\mathrm{c}} \right) \right] \simeq  39.1  ^{\circ}$ with $n=1.52$ the refractive index of the cover glass. $\theta_c=\arcsin \left( \frac{1}{n}\right)$ is the incident angle enabling the coupling between the light in the cover glass and the surface plasmon propagating on the interface upper electrode/vacuum. Figure \ref{fig:Kretchmann_Hfield} shows the amplitude of the normalized magnetic field $|H/H_0|$ inside the junction.  The measured SPR (see Fig.~\ref{fig:PAT_angle_300K}) is broader than the simulated one probably due to the roughness of our ultra-thin film deposition. At angle $\alpha_0$, the field enhancement occurs on the upper electrode. The electric field is deduced from the magnetic field:

\begin{equation}\label{eq.Efield}
 \mathbf{E_i}= -\frac{H_i}{\omega \epsilon_0 \epsilon_i}\left( k_z \mathbf{\mathbf{e_\mathrm{x}}} +k_x \mathbf{\mathbf{e_\mathrm{z}}}\right).
\end{equation}

\begin{figure}[htbp!]
\begin{center}
\includegraphics[width=7cm]{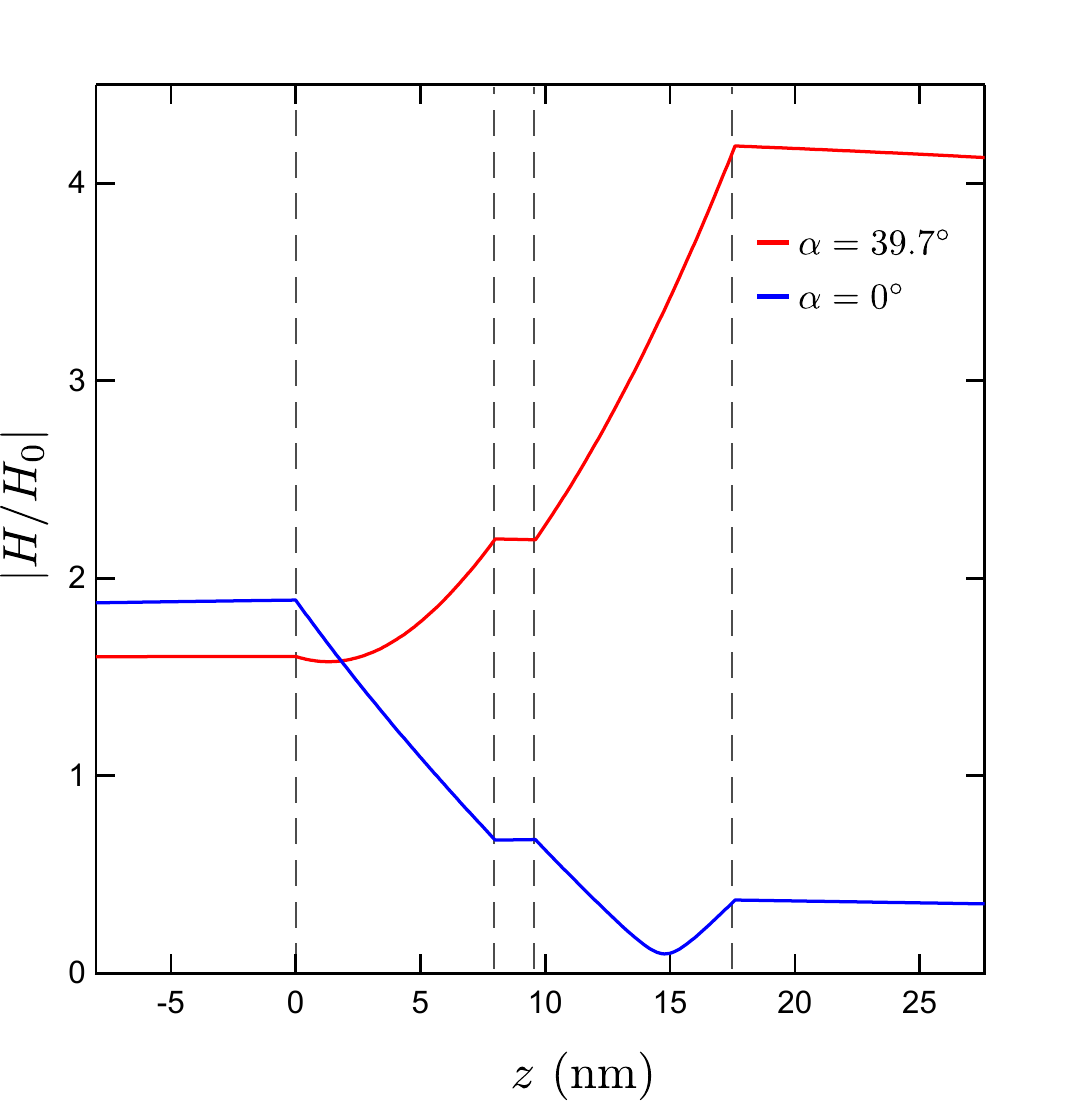}
\end{center}
\caption{Magnetic field as a function of $z$  at normal incidence $\alpha=0 ^{\circ}$ and at SPR $\alpha \simeq 39.7 ^{\circ}$ \label{fig:Kretchmann_Hfield_z_alpha}}
\end{figure}

\begin{figure}[htbp!]
\begin{center}
\includegraphics[width=7cm]{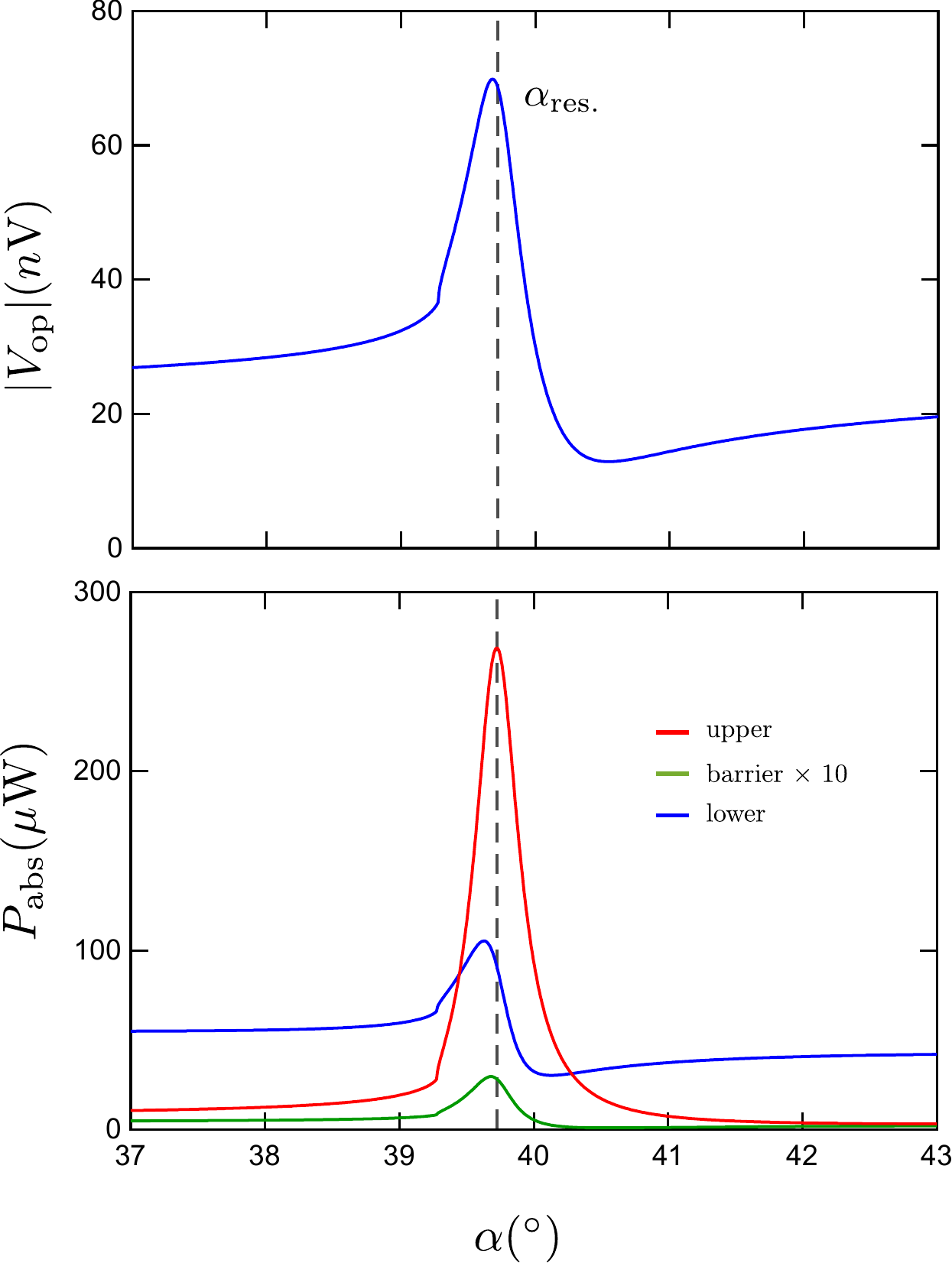}
\end{center}
\caption{\textbf{a}. Optical voltage between the two electrode as a function of angle $\alpha$ for $P_{\mathrm{laser}}=53 \, \mathrm{mW}$. \textbf{b}. Absorbed power in the tunnel function  as a function of angle $\alpha$ for $P_{\mathrm{laser}}=53 \, \mathrm{mW}$. \label{fig:Kretchmann_Vop_P}}
\end{figure}

\noindent  The optical voltage between the two upper and lower electrodes is then given by:

\begin{equation}\label{eq.Vop}
 V_{\mathrm{op}}(\alpha)= \left| \int_{a}^{a+d} \mathbf{E_i}.\mathbf{\mathbf{e_\mathrm{z}}} \, dz \right|= \left| \frac{k_x}{\epsilon_0 \epsilon_i \omega}   \right| . \left| \int_{a}^{a+d} H_i(z,\alpha) \, dz \right| .
\end{equation}

\noindent and it is maximum around the SPR: $V_{\mathrm{op}}(\alpha_0)\simeq 25 \, \mu \mathrm{V} $ (see Fig.~\ref{fig:Kretchmann_Vop_P} \textbf{a}). The absorbed power in the different regions of the junction is given by:

\begin{equation}\label{eq.Pop}
 P_{\mathrm{\mathrm{abs.},i}}(\alpha)= \frac{\omega \mathrm{Im} \left( \epsilon_i\right)}{2 \epsilon_0 \left| \epsilon_i\right|} \int_{{i}} \left|H_i(z,\alpha) \right|^2 \, dz.
\end{equation}

\noindent Figure \ref{fig:Kretchmann_Vop_P}\textbf{b} shows that the upper electrode absorbs more power than the lower one at SPR: $P_{\mathrm{up}}\simeq 79 \, \mathrm{\mu W}$ whereas $P_{\mathrm{dn}}\simeq 24 \, \mathrm{\mu W}$. Note that the power dissipated in the barrier is negligible and that it is reasonable to assume that the barrier transmission is not affected by any thermal effects.

\section*{Appendix B: Tunnel transmission}

\begin{figure}[htbp!]
\begin{center}
\includegraphics[width=7cm]{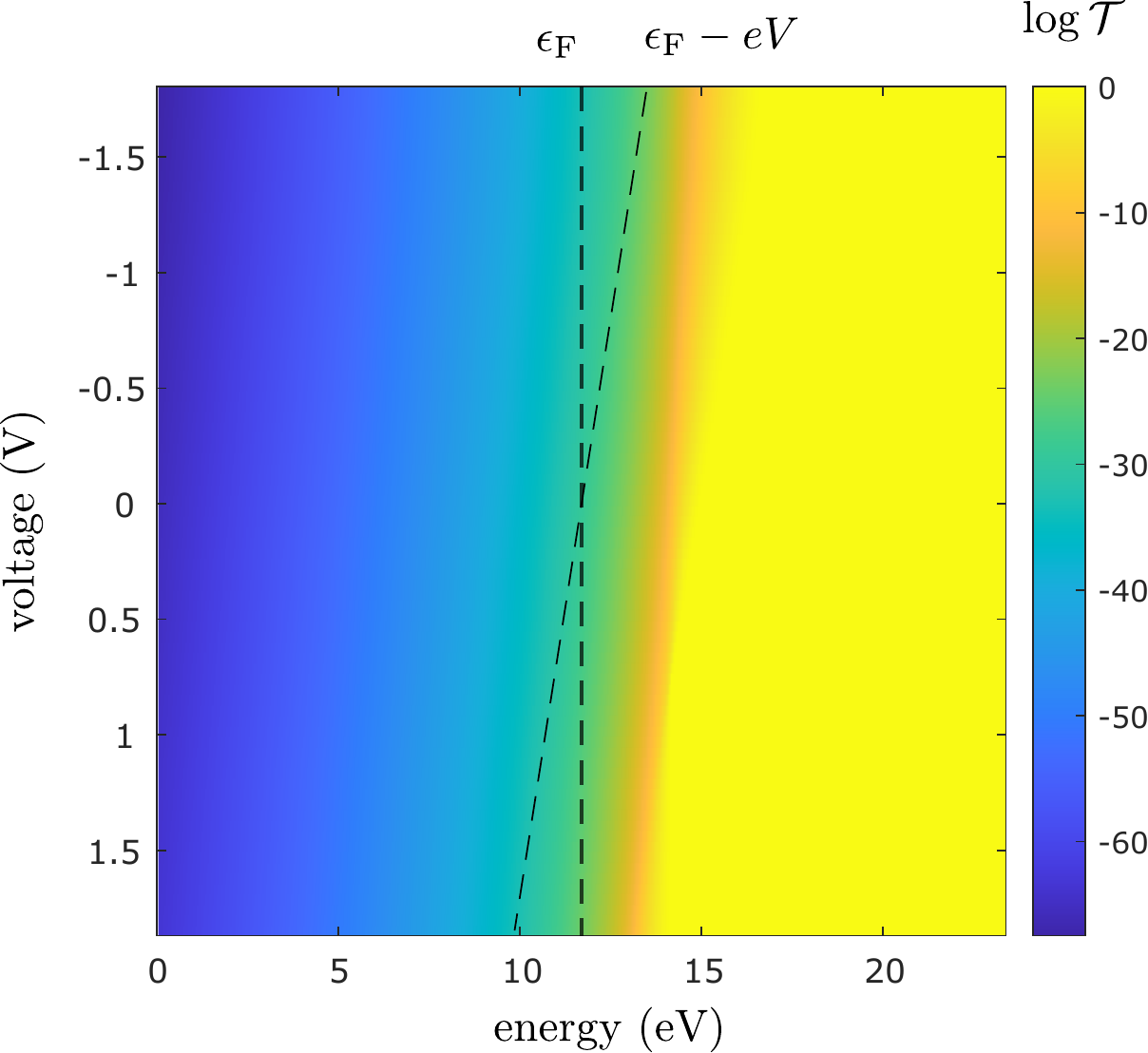}
\end{center}
\caption{Transmission $\mathcal{T}(\epsilon,eV_{\mathrm{dc}})$ for a trapezoidal barrier characterized by $U_0=2.87 \, \mathrm{eV}$, $\Delta U =-0.78 \, \mathrm{eV}$ and $d=1.6 \, \mathrm{nm}$. \label{fig:Transmission}}
\end{figure}

\begin{figure}[htbp!]
\begin{center}
\includegraphics[width=4cm]{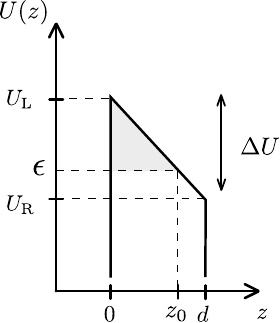}
\end{center}
\caption{Trapezoidal barrier defined by de thickness $d$, a mean high $U_0$ and an assymetry $\Delta U$ . \label{fig:barrier}}
\end{figure}

\noindent In order to calculate the different quantities related to the photon-assisted tunneling current in the far from equilibrium regime, we model the tunnel barrier by a trapezoidal barrier $U(z)$ characterized by a thickness $d$, a mean height $U_0$ and an asymmetry $\Delta U$ (see Fig.~\ref{fig:barrier}). This allowed us to obtain an analytical expression of the tunnel transmission $\mathcal{T}(\epsilon,eV_{\mathrm{dc}})$. This analytical expression is essential to perform numerical optimization on the $I(V)$ characteristics (see Fig.~\ref{fig:PATI_volt_pwr}) and the photon assisted current  (see Fig.~\ref{fig:PATI_current_fit}  and eq.~(\ref{eq.LB3})).

\noindent  We are considering  the Simmons' model where an effective potential includes the potential barrier $U(z)$ and the biasing energy $U_{\mathrm{bias}}(z,eV)=-eV(1-z/d)$, $U_{\mathrm{eff}}(z)=U_{\mathrm{bias}}(z,eV)=U(z)+U_{\mathrm{bias}}(z,eV)$. The 3D transmission can be expressed as a function of the Wentzel–Kramers–Brillouin  (WKB) transmission through the 1D effective potential \cite{Simmons_1964,Fevrier_2018}:

\begin{equation}\label{eq.WKB}
\mathcal{T}= \exp \left\{-\frac{\sqrt{8m}}{\hbar} \int_{0,z_0}^{z_0,d} \,dz \sqrt{U(z)-eV\left(1-\frac{z}{d}\right)-\epsilon} \, \right\}.
\end{equation}

\noindent We can distinguish three cases: (i) the energy $\epsilon$ is smaller that the barrier and the electrons see a trapezoidale barrier ($\epsilon < \min_{0<z<d}{U(z)}$) (ii) the electrons see a triangular barrier ($\min_{0<z<d}{U(z)}  < \epsilon < \max_{0<z<d}{U(z)}$ and (iii) the electrons go over the barrier ($\epsilon > \max_{0<z<d}{U(z)}$). These cases depend on the values of $\epsilon$ and $eV$:

\begin{itemize}
  \item if $\epsilon < \mathrm{min} \left( U_{\mathrm{L}}-eV,  U_{\mathrm{R}} \right)$ then
  $$\log \mathcal{T}=-\frac{4d\sqrt{2me}}{3} \frac{\left(U_{\mathrm{R}}-\epsilon \right)^{3/2}-\left(U_{\mathrm{L}}-eV-\epsilon \right)^{3/2}}{U_{\mathrm{R}}-U_{\mathrm{L}}+eV},$$
  \item if $ U_{\mathrm{L}}- eV < \epsilon < U_{\mathrm{R}}$ then
  $$\log \mathcal{T}=-\frac{4d\sqrt{2me}}{3} \frac{\left(U_{\mathrm{R}}-\epsilon \right)^{3/2}}{U_{\mathrm{R}}-U_{\mathrm{L}}+eV},$$
  \item if $ U_{\mathrm{R}} < \epsilon< U_{\mathrm{L}}-eV$ then
  $$ \log \mathcal{T}=-\frac{4d\sqrt{2me}}{3} \frac{\left(U_{\mathrm{L}}-eV-\epsilon \right)^{3/2}}{U_{\mathrm{L}}-U_{\mathrm{R}}-eV},$$
  \item if $  \epsilon > \mathrm{max} \left( U_{\mathrm{L}}-eV,  U_{\mathrm{R}} \right)$ then
  $$\mathcal{T}=1.$$
\end{itemize}

\noindent The Landauer-Büttiker formula gives the $I(V)$ characteristics:
\begin{equation}\label{eq.Buttiker}
I(V)=\frac{2e}{h}\int d\epsilon \, \mathcal{T}(\epsilon,eV) \left[ f(\epsilon)-f(\epsilon+eV) \right].
\end{equation}
\noindent A numerical optimization fitting of the $I(V)$ characteristics with eq.~(\ref{eq.Buttiker}) and the analytical transmission gives $U_0=2.87 \, \mathrm{eV}$, $\Delta U =-0.78 \, \mathrm{eV}$ and $d=1.6 \, \mathrm{nm}$. Fig.~\ref{fig:Transmission} shows the color plot of $\mathcal{T} \left(\epsilon,V_{\mathrm{dc}}\right)$ for the considered barrier.
\section*{Appendix C: Non-equilibrium distribution function}

\begin{figure}[H]
\vspace{0.5cm}
\begin{center}
\includegraphics[width=7cm]{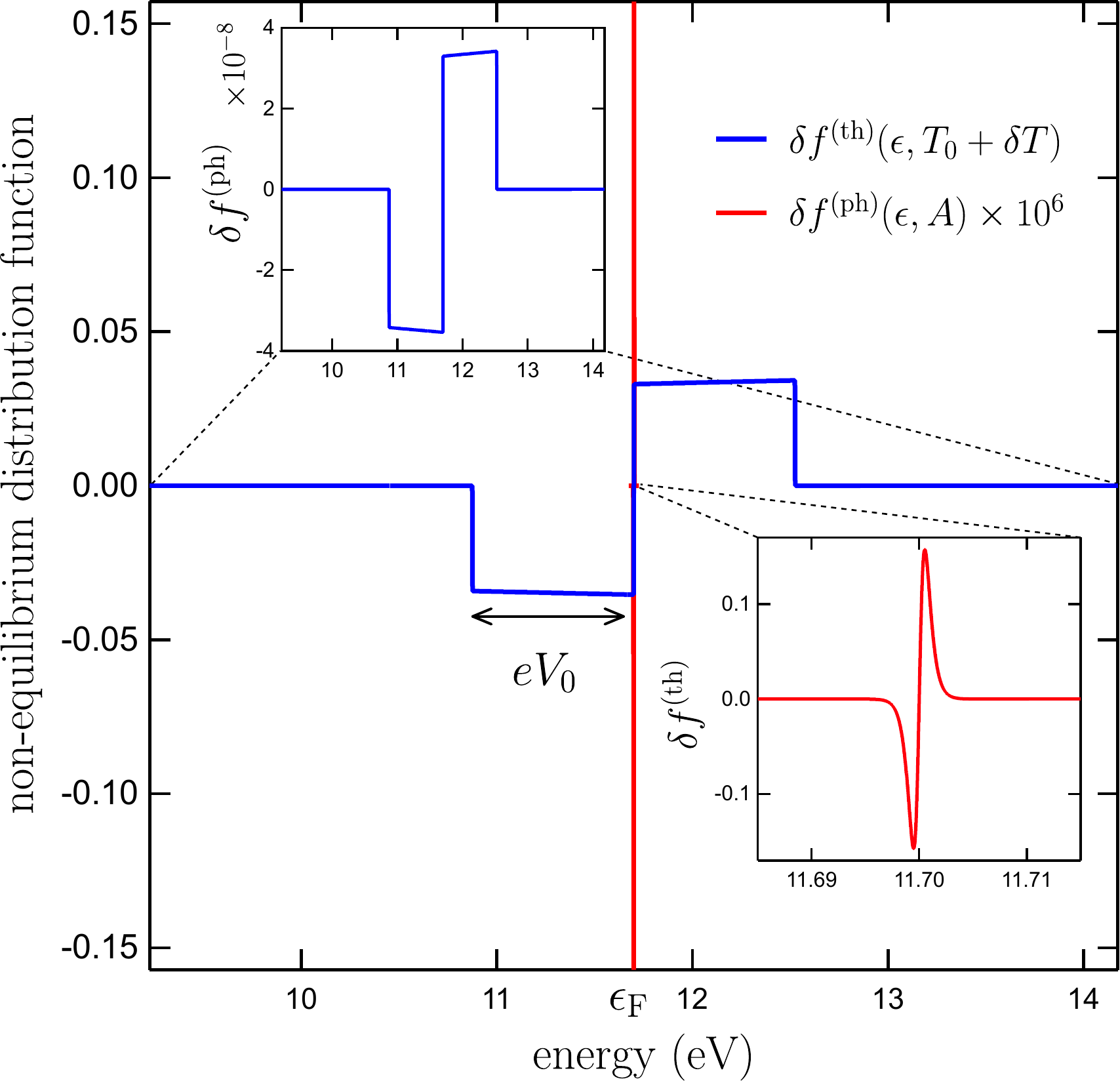}
\end{center}
\caption{Non-equilibrium distribution function due to photon-assisted processes/heating. The settings parameters correspond to the measured one in section V: $\left( T_0=2.8 \, \mathrm{K}, \delta T=3,\ \mathrm{K} \right)$  for heating and $A_{\mathrm{exc}}=10^{-8}$ for photon illumination. \label{fig:distrib}}
\end{figure}

\noindent The interplay between photons and electrons in the metallic contact may lead to "hot" electron distribution junction or heating. For a moderate increase of temperature, the heating affects a large number of quasiparticles around the Fermi sea whereas the photon-assisted processes brings a small amount of quasiparticles at high energy corresponding to the photon energy $\hbar \omega$ (see Fig. \ref{fig:distrib}). The excitation of the conduction electrons is done by the absorption of  photons of energy $\hbar \omega$. Assuming that the processes of excitation involve only intraband transitions, electrons of energy $\epsilon$ reach  energy $\epsilon +\hbar \omega$. The change in occupancy number $\delta f_{\mathrm{exc}}$ is given by the difference between the number $dN_{+}(\epsilon)$ of electrons that access the energy $\epsilon$ and the number $dN_{-}(\epsilon)$ of electrons that leave it:

\begin{subequations}
\begin{align*}
dN_{+}(\epsilon)&=A \rho(\epsilon-\hbar \omega) f_0(\epsilon-\hbar \omega) \rho(\epsilon)\left( 1-f_0(\epsilon)\right),\\
dN_{-}(\epsilon)&=A \rho(\epsilon) f_0(\epsilon) \rho(\epsilon+\hbar \omega)\left( 1-f_0(\epsilon+\hbar \omega)\right),\\
\nonumber
\end{align*}
\end{subequations}

\noindent where $\rho(\epsilon) \propto \sqrt{\epsilon}$ is the density of states of the free electron gas, $A$ is a constant proportional to the energy transfered to the electron gas and $f_0$ the equilibrium Fermi-Dirac distribution. The change in the occupancy number is  given by $\delta f_{\mathrm{exc}}=\left( dN_{+}(\epsilon)-dN_{-}(\epsilon)\right)/\rho(\epsilon)$. The excitation term in eq.~(\ref{eq.distrib_ph_abs}) is then directly deduced.

\begin{figure}[H]
\vspace{0.5cm}
\begin{center}
\includegraphics[width=7cm]{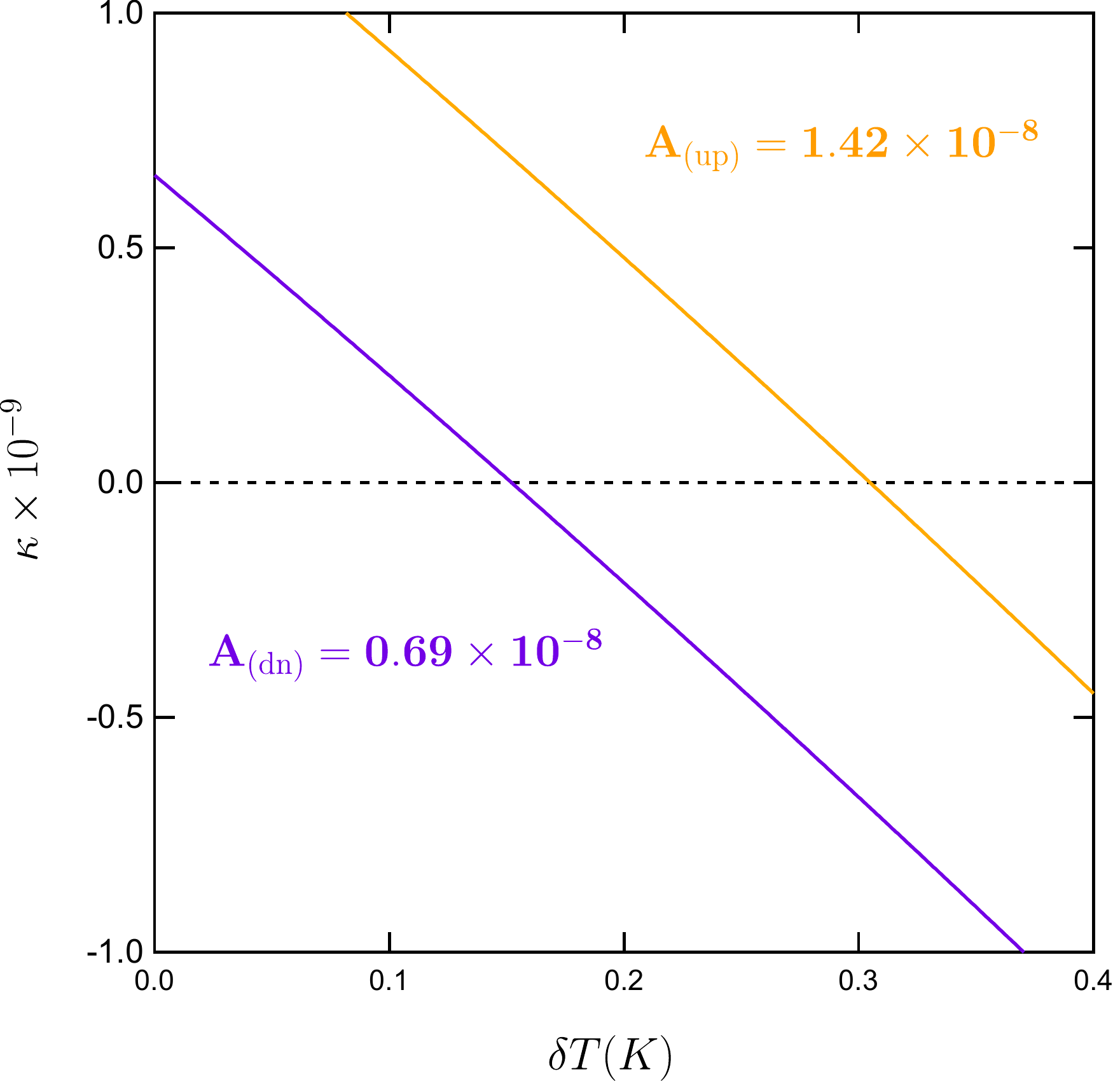}
\end{center}
\caption{Normalized difference $\eta(A,\delta T)$ for the equivalent temperature of excitation in the upper electrode characterized by a power coefficient $A_{\mathrm{up}}=1.42 \times  10^{-8}$ (orange line) and the lower electrode characterized by a power coefficient $A_{\mathrm{dn}}=0.69 \times 10^{-8}$ (purple line).  \label{fig:kappa}}
\end{figure}

\noindent  The normalized difference between the energy contained in the non-equilibrium distribution function characterized by absorbed power coefficient $A$ and the equilibrium function at temperature $T_0+\delta T$ is given by:

\begin{equation*}
\kappa(A,\delta T)=\frac{\int \epsilon \rho(\epsilon) \left[ f(\epsilon,T_0)-f(\epsilon,T_0+\delta T)+\delta f_{\mathrm{exc}} \right]\, d\epsilon}{\int \epsilon \rho(\epsilon) f(\epsilon,T_0) \, d\epsilon}.
\end{equation*}

\noindent The equivalent temperature of excitation $\delta T_{\mathrm{exc}}$ corresponds to $\kappa(A_{\mathrm{exc}},\delta T_{\mathrm{exc}})=0$. Figure \ref{fig:kappa} gives the values of $\delta T_{\mathrm{exc}}$ for the upper electrode ($A_{\mathrm{up}}= 1.42 \times 10^{-8}$) and the lower one ($A_{\mathrm{dn}}= 0.69 \times 10^{-8}$).

\section*{Appendix D: Temperature dependence of the conductivity}

\noindent Using the free electron gas model to describe the electron transport in the metallic electrodes, the Boltzmann equation and the relaxation time approximation give the electrical conductivity:
\begin{equation}\label{eq.conductivity}
\sigma= \frac{ne^2\langle \tau \rangle}{m^{\star}},
\end{equation}
\noindent where $n$ is the density of electrons, $m^{\star}$ the effective masse of electrons, $e$ the elementary charge and $\langle \tau \rangle$ the mean value of the relaxation time $\tau(\epsilon)$. By considering the density of states $\rho \propto \epsilon^{3/2}$ in 3D, the mean value is given by:
\begin{equation}\label{eq.tau}
\langle \tau \rangle= \frac{\int_{0}^{+ \infty} \tau(\epsilon)\epsilon^{3/2} \left(-\frac{\partial f_0}{\partial \epsilon} \right) d\epsilon}{\int_{0}^{+ \infty} \epsilon^{3/2} \left(-\frac{\partial f_0}{\partial \epsilon} \right) d\epsilon},
\end{equation}
\noindent In the first order approximation near the Fermi energy, the relaxation time can be written as $\tau(\epsilon) \simeq \tau(\epsilon_{\mathrm{F}})+ \tau'(\epsilon_{\mathrm{F}}) \left( \epsilon-\epsilon_{\mathrm{F}} \right)$. The Sommerfeld expansion of $\langle \tau \rangle$  gives:
\begin{equation}\label{eq.sommerfeld_tau}
\langle \tau \rangle=\tau(\epsilon_{\mathrm{F}})+ \frac{\pi^2}{2} \frac{\tau'(\epsilon_{\mathrm{F}})}{\epsilon_{\mathrm{F}}} \left( k_{\mathrm{B}}T\right)^2,
\end{equation}
\noindent and the relative increase of the resistance is thus:
\begin{equation}\label{eq.eta}
\eta_{\mathrm{up/dn}}=\frac{\delta R_{\mathrm{up/dn}}}{R_{\mathrm{up/dn}}}=-\frac{\pi^2\tau'(\epsilon_{\mathrm{F}})}{2\epsilon_{\mathrm{F}}\tau(\epsilon_{\mathrm{F}})}\left( k_{\mathrm{B}}\delta T_{\mathrm{up/dn}}\right)^2.
\end{equation}
\noindent Assuming that the energy dependence of the relaxation time is the same in both aluminum electrodes, the ratio $\eta_{\mathrm{up}}/\eta_{\mathrm{dn}}$ gives access to the ratio $\delta T_{\mathrm{up}}/\delta T_{\mathrm{dn}}$.

\vspace{0.2cm}\noindent\emph{Acknowledgements.}
We acknowledge fruitful discussions with Marc Bescond. We also thank Sylvie Gautier for the help in sample fabrication and Gilles Guillier for the cryognetic rotation realization. This work has been supported by Region Ile-de-France in the framework of DIM SIRTEQ.

\end{document}